\documentstyle[11pt]{article}

\evensidemargin=\oddsidemargin
\begin{document}

\begin{flushright}
{\large \sf  TUIMP-TH-94/58\\}
\end{flushright}
\bigskip
\begin{center}
{\Large \bf CALCULATION OF THE FOUR-QUARK CONDENSATES IN THE QCD MOTIVATED
            EXTENDED NAMBU-JONA-LASINIO MODEL}
\footnote{ Work supported by the National Natural Science Foundation of
China and the Fundamental Research Foundation of Tsinghua University}
\\[1.0cm]
{\bf Qing Wang\hspace{0.8cm} Yu-Ping Kuang\hspace{0.8cm} Yu-Ping Yi} \\[0.1cm]
China Center of Advanced Science and Technology(World Laboratory), P.O.Box
8730, Beijing 100080, China; and Institute of Modern Physics, Tsinghua
University, Beijing 100084, China\footnote{Mailing address}; and Institute of
Theoretical Physics, Academia Sinica, Beijing  100080, China \\[0.2cm]
and\\[0.2cm]
{\bf Tian-Fang Cai}\\[0.1cm]
Department of Physics, Tsinghua University, Beijing 100084, China
\end{center}

\vspace{1.8cm}
\centerline{\bf Abstract}
\vspace{0.2cm}
\begin{sf}

The four-quark condensates appearing in QCD sum rule are calculated
nonperturbatively in a realistic QCD motivated extended Nambu-Jona-Lasinio
model. The calculation is in the framework of the effective potential for local
composite operators up to next-to-the-leading order in the 1/N expansion in
which the non-factorized parts of the condensates are included. We show in
this paper the possibility of explaining the phenomenological enhancement
factor $~\kappa=2.4~$, needed for fitting the data in the factorization
approximation, as the contribution of the non-factorized parts of the four-
quark condensates.

\vspace{0.4cm}
\end{sf}

\newpage
%\vspace{2.0cm}
\begin{sf}
%\noindent
%\null
One of the interesting but difficult problems in nonperturbative QCD is
the calculation of the quark and gluon condensates. The two-quark condensate
breaks the chiral symmetry which explains the lightness of the low lying
$~0^-$ mesons, while all condensates are relevant to the well-known QCD sum
rule \cite{QCDsr}. QCD sum rule incorporates the quark and gluon condensates
with perturbative QCD calculations. For example, in the study of the sum rule
for mesons, the vacuum expectation value (VEV) of two current operators
$~<Tj^A(x)j^B(0)>~$ is considered, and the operator product expansion (OPE)
gives
\begin{eqnarray}                       %(1)
i\int d^4x\,e^{iq \cdot x}\,<Tj^A(x)j^B(0)>\,=\,\sum_n\,C^{AB}_n(q)\,<O_n>~,
\end{eqnarray}
\noindent
where $~C^{AB}_n(q)~$'s are perturbatively calculable Wilson coefficients and
$~<O_n>~$'s are various nonperturbative vacuum condensates. In practical
applications, on the RHS of (1), only $~C^{AB}_n(q)~$'s are calculated from
perturbative QCD while  the condensates $~<O_n>~$'s are left as free
parameters determined by experimental inputs. To reduce the number of free
parameters, people often make the simple {\it factorization} approximation
to express the four-quark condensate in terms of the two-quark condensate,
i.e.
\begin{eqnarray}         %(2)
<{\bar \psi}\Gamma\psi{\bar \psi}\Gamma\psi>\,\approx~ F_{\Gamma}<{\bar \psi}
\psi>^2~,
\end{eqnarray}
\noindent
where the Lorentz structure of $~\Gamma~$ is $~1~$ for scalar (S), $~i\gamma_5
{}~$ for pseudoscalar (P), $~\gamma_{\mu}~$ for vector (V), $~\gamma_5\gamma_
{\mu}~$ for axial-vector (A), $~\sigma_{\mu\nu}~$ for tensor (T) (for color
and flavor non-singlet, the color and flavor group generators $~\lambda_\alpha
/2~$, $~t_i~$ should also be included ), and $~F_{\Gamma}~$ is a constant
depending on the structure of $~\Gamma~$ and can be easily calculated
\cite{QCDsr}. However, applications of QCD sum rule to various hadronic
processes with the approximation (2), e.g. the sum rules for baryons \cite{D}
$\rho$-meson \cite{LNT} and pseudoscalar meson  \cite{KPS}, show that the
theoretical results are not quite good and phenomenologically the $~F_{\Gamma}
<{\bar \psi}\psi>^2~$ term should be enlarged by a factor $~\kappa\approx
3.6~$ for fitting the data \cite{D}\cite{LNT}\cite{KPS}. Further study of the
perturbative QCD corrections to the Wilson coefficient of the operator $~({
\bar \psi}\Gamma\psi)^2~$ shows that the $~O(\alpha_s)~$ correction increases
the Wilson coefficient by a factor of $~1.5~$ \cite{LSC} which means that the
enhancement factor should actually be $~\kappa\approx 2.4~$.

So far there is no theoretical explanation of the enhancement factor $~\kappa
\approx 2.4~$. A possible and natural conjecture is that the need for an
enhancement factor might be due to the neglect of the non-factorized part of
the four-quark condensate $~<{\bar \psi}\Gamma\psi{\bar \psi}\Gamma\psi>_{NF}~$
 in (2), but this is difficult to prove from the first principles of QCD.
Several attempts have been made to discuss this problem. A careful calculation
of the anomalous dimensions of the four-quark operators shows that the
deviation from (2) varies significantly with the renormalization scale, i.e.
even if (2) is good at low energy it may fail at some high energy scale \cite
{JK}. In Ref.\cite{HK}, a general formalism of the effective potential for
local composite operators in the framework of $~1/N~$ expansion was developed
and was applied to calculate the four-fermion condensate $~<{\bar \psi}\psi
{\bar \psi}\psi>~$ nonperturbatively in the Gross-Neveu model. It is shown that
 (2) {\it holds only in the $~N \rightarrow \infty~$ limit}, i.e. $~<{\bar\psi}
\psi{\bar \psi}\psi>_{NF}~$ {\it is nonvanishing to order-$1/N$}. Up to order-
$1/N$, for $~N=3~$, the numerical result shows that the ratio $~R\equiv <{\bar
\psi}\psi{\bar\psi}\psi>_{NF}/[F_S<{\bar \psi}\psi>^2]~$ runs rapidly with the
renormalization scale, and the minimal value of $~R~$ is about $~1/3~$. This
means that (2) is not a good approximation even at low energy. However, the
Gross-Neveu model is a toy model which can not predict the value of $~\kappa~$
to compare with the experiment. In recent years, there have been several papers
 studying QCD motivated Nambu-Jona-Lasinio (NJL) models as effective theories
for low lying hadrons\cite{BBdR}. These models lead to quite successful
phenomenological predictions, especially the extended Nambu-Jona-Lasinio (ENJL)
 model in Ref. \cite{BBdR}, so that they may reflect some main features of
nonperturbative QCD. Since the NJL model is a four-fermion interaction theory
similar to the Gross-Neveu model, the method developed in Ref.\cite{HK} can be
directly applied to it. In this paper, we take the EJNL model of Ref.\cite
{BBdR} and calculate the four-quark condensates appearing in the QCD sum rule
nonperturbatively by using the method of Ref.\cite{HK}. The four-quark
condensates are generally related to the sum rule in the form $~\displaystyle
\sum_{\Gamma}C_6^{\Gamma}<{\bar \psi}\Gamma\psi{\bar \psi}\Gamma\psi>~$ (cf.
(1)). Once the condensates $~<{\bar \psi}\psi>~$ and $~<{\bar \psi}\Gamma\psi
{\bar \psi}\Gamma\psi>~$ are calculated, the enhancement factor $~\kappa~$ can
be obtained from
\begin{eqnarray}             %(3)
\kappa\,=\,\displaystyle{ {\frac {\displaystyle\sum_{\Gamma}\,C_6^{\Gamma}
<{\bar \psi}\Gamma\psi{\bar \psi}\Gamma\psi>}{\displaystyle\sum_{\Gamma}
C_6^{\Gamma}F_{\Gamma}<{\bar \psi}\psi>^2}}}\,
=\,1\,+\,\displaystyle{ {\frac{\displaystyle\sum_{\Gamma}\,C_6^{\Gamma}
<{\bar \psi}\Gamma\psi{\bar \psi}\Gamma\psi>_{NF}}{\displaystyle\sum_{\Gamma}
C_6^{\Gamma}F_{\Gamma}<{\bar \psi}\psi>^2}}}~~.
\end{eqnarray}

Here we sketch the main points of the method developed in Ref.\cite{HK}. To
calculate the two-quark and four-quark condensates, we consider the following
generating functional
\begin{equation}              %(4)
\begin{array}{ll}
Z[K_2,K_4]\,\equiv expiW[K_2,K_4]\,=\,i\displaystyle{\int {\cal D}\psi
{\cal D}{\bar \psi}\, expi\int d^4x\,\{\,{\cal L}\,+\,\displaystyle{\sum_
{\Gamma}K^{\Gamma}_2{\bar \psi}\Gamma\psi\,+\,\sum_{\Gamma}K^{\Gamma}_4{\bar
\psi}\Gamma\psi{\bar \psi}\Gamma\psi}}~\\\;\;\;\;\;\;\;\;\;\;\;\;\;\;\;\;\;\;\;
\;\;\;+\,P(K_2,K_4)\}~,
\end{array}
\end{equation}
\noindent
where $~K^{\Gamma}_2~$ and $~K^{\Gamma}_4~$ are local external sources, and
$~P(K_2,K_4)~$ is a pure source term needed for satisfying the consistency
condition requiring that no condensate effect occurs at the classical level
\cite{KLZ}. Let $~W_c~$ be the classical value of $~W~$. The consistency
condition is \cite{KLZ}
\begin{equation}             %(5)
\begin{array}{ll}
\displaystyle{\left.\frac{\delta^{n+m}W_c}{\delta (K^{\Gamma}_2)^n \delta
(K^{\Gamma}_4)^m}\,\right |}_{K_2,K_4=0}\,=\,0~~.
\end{array}
\end{equation}
\noindent
Here, for simplicity, we have ignored the external sources for the single
fields $~\psi~$ and $~\bar \psi~$ in (4) since we are not calculating
$~<\psi>~$ and $~<{\bar \psi}>~$ which actually vanish. Define the classical
fields $~\Sigma_{\Gamma}~$ and $~\Xi_{\Gamma}~$,
\begin{equation}             %(6)
\begin{array}{ll}
\displaystyle{\frac {\delta W}{\delta K^{\Gamma}_2}}\,=\,\Sigma_{\Gamma}\,~,
\;\;\;\;\;\;\;\;\displaystyle{\frac {\delta W}{\delta K^{\Gamma}_4}}\,=
\,\Sigma^2_{\Gamma}\,+\,\,\Xi_{\Gamma}~~,
\end{array}
\end{equation}
\noindent
$~\Sigma_S~$ and $~\Xi_{\Gamma}+(1-F_{\Gamma})\Sigma_S^2~$ give $~<{\bar \psi}
\psi>~$ and $~<{\bar \psi}\Gamma\psi{\bar \psi}\Gamma\psi>_{NF}~$,
respectively, in the vacuum state. Once $~W[K_2,K_4]~$ is obtained, we can
make the Legendre transformation
\begin{equation}            %(7)
\Gamma[\Sigma,\Xi]\,=\,W[K_2,K_4]\,-\,\displaystyle{\int d^4x\,\{\,\sum_
{\Gamma}K^{\Gamma}_2\Sigma_{\Gamma}\,+\,\sum_{\Gamma}K^{\Gamma}_4\,(\,
\Sigma^2_{\Gamma}\,+\,\Xi_{\Gamma}\,)\,\}}~,\\
\end{equation}
\noindent
to get the effective action $~\Gamma[\Sigma,\Xi]~$ and so the effective
potential $~V_{eff}(\Sigma,\Xi)~$. The rules for calculating $~W[K_2,
K_4]~$ are as follows. Define the well-known propagator \cite{J}
\begin{equation}             %(8)
\begin{array}{ll}
i{\cal D}^{-1}(x,y;K_2)\,=\,\displaystyle{\left.\frac{\delta^2S}{\delta \psi(y)
\delta {\bar \psi}(x)}\,\right |}_{\psi,{\bar \psi}=0~},
\end{array}
\end{equation}
\noindent
where $~S~$ is the action $~S=\displaystyle{\int d^4x\,[\,{\cal L}\,+\,
\sum_{\Gamma}K^{\Gamma}_2{\bar \psi}\Gamma\psi\,+\,\sum_{\Gamma}K^{\Gamma}_4
{\bar \psi}\Gamma\psi{\bar \psi}\Gamma\psi\,+\,P(K_2,K_4)]}~$. $~{\cal D}~$ is
related to the physicsl propagator $~G~$ by
\begin{equation}           %(9)
G^{-1}\,=\,{\cal D}^{-1}\,-\,\Pi~,
\end{equation}
\noindent
where $~\Pi~$ is the proper self-energy of the $~\psi~$ field. For models like
the ENJL model, we can always consider a special proper self-energy diagram
$~\Pi_s~$, composed of physical propagators and bare vertices, which is
momentum-independent and is of the following structure \cite{HK}
\begin{eqnarray}          %(10)
\Pi_s\,=\,-i\displaystyle{\sum_{\Gamma}b_{\Gamma}\Gamma\Delta_{\Gamma}}~,
\end{eqnarray}
\noindent
where
\begin{eqnarray}          %(11)
\Delta_\Gamma\,\equiv \,-\,{\frac {\delta W_L}{\delta K^{\Gamma}_2}}\,
=\,Tr(\Gamma G)~,
\end{eqnarray}
\noindent
and $~b_{\Gamma}~$ is a constant determined by the model of interaction. In
(11) $~W_L~$ means the loop contribution to $~W~$. Define a new propagator
$~G_s~$ ,
\begin{eqnarray}         %(12)
G_s^{-1}\,=\,{\cal D}^{-1}\,-\,\Pi_s~.
\end{eqnarray}
\noindent
In terms of these notations, the formula for $~W[K_2,K_4]~$ derived in Ref.
\cite{HK} is written as
\begin{equation}        %(13)
\begin{array}{ll}
W[K_2,K_4]\,=\,\displaystyle{\int d^4x\,\{\,{\cal L}\,+\,\displaystyle{\sum_
{\Gamma}K^{\Gamma}_2{\bar \psi}\Gamma\psi\,+\,\sum_{\Gamma}K^{\Gamma}_4{\bar
\psi}\Gamma\psi{\bar \psi}\Gamma\psi}\,+\,P(K_2,K_4)\}\,-\,i\,TrlniG_s^{-1}}\\
\;\;\;\;\;\;\;\;\;\;\;\;\;\;\;\;\;\;\;\;\;\;\;-\,\displaystyle{\int d^4x\,\sum_
{\Gamma}b_{\Gamma}\Delta_{\Gamma}^2}\,-\,i<0|expi\int d^4x\,{\cal L}_I|0>^
\prime_{P2PI(\Pi_s)}
{}~,
\end{array}
\end{equation}
\noindent
where $~{\cal L}_I~$ is the interaction Lagrangian, and the last term means
the sum of all the partially two-particle irreducible diagrams with respect to
$~\Pi_s~$ (with propagators $~G_s~$'s and vertices in $~{\cal L}_I~$) defined
in Ref. \cite{HK} except the type of diagrams shown in Fig.(4c) in Ref. \cite
{HK}. The first term in (13) is the tree level contribution, while other terms
are contributed by loop diagrams ( $~W_L~$). In (13), $~\Delta~$ serves as an
unknown constant which can be further solved from (11). Then the effective
potential $~V_{eff}~$ can be obtained from (13) and (7). The condensates can
then be determined by the minimum of $~V_{eff}~$.

Actually, once $~W~$ is obtained, we can also calculate the condensates simply
from (6) by taking $~K_2=K_4=0~$ without knowing $~V_{eff}~$. The solutions of
(6) are the same as those obtained from the extremum condition of $~V_{eff}~$
as it should be\cite{HK}. To pick up the true condensates (minimum of $~V_{eff}
{}~$), one needs only calculate the values of $~V_{eff}~$ for these solutions.
This is much easier than calculating the complete function $~V_{eff}(\Sigma,
\Xi)~$ for arbitrary $~\Sigma~$ and $~\Xi~$.

The Lagrangian in the ENJL model in Ref. \cite{BBdR} is
\begin{eqnarray}            %(14)
{\cal L}_{QCD}\,=\,{\cal L}^{\Lambda_\chi}_{QCD}\,+\,{\cal L}^{SP}_{NJL}~
+\,{\cal L}^{VA}_{NJL}~,
\end{eqnarray}
\noindent
where $~\Lambda_\chi~$ is a momentum cut-off, $~{\cal L}^{\Lambda_\chi}_
{QCD}~$ is the $~QCD~$ Lagrangian for momentum below $~\Lambda_\chi~$, and
\begin{equation}        %(15)
\begin{array}{ll}
{\cal L}^{SP}_{NJL}\,=\,\displaystyle{{\frac {8\pi^2G_S}{N_c\Lambda^2_\chi}}
\sum_{ab}({\bar \psi}^a_R\psi^b_L)({\bar \psi}^b_L\psi^a_R)}~\\~\;\;\;
\;\;\;\;\;\;\,=\,\displaystyle{\frac{2\pi^2G_S}{N_cN_f\Lambda^2_\chi}[({\bar
\psi}\psi)^2\,+\,({\bar \psi}i\gamma_5\psi)^2]}~\\~\;\;\;\;\;\;\;\;\;\;\;\;\;\,
+\,\displaystyle{{\frac {2\pi^2{\hat G}_S}{N_cN_f\Lambda^2_\chi}}
\sum_{i=1}^{N^2_f-1}[({\bar \psi}t_i\psi)({\bar \psi}t_i\psi)\,+\,({\bar \psi}
i\gamma_5t_i\psi)({\bar \psi}i\gamma_5t_i\psi)]}~,
\end{array}
\end{equation}

\begin{equation}       %(16)
\begin{array}{ll}
{\cal L}^{VA}_{NJL}\,=\,-\,\displaystyle{{\frac {8\pi^2G_V}{N_c\Lambda^2_\chi}}
\sum_{ab}[({\bar \psi}^a_L\gamma^{\mu}\psi^b_L)({\bar \psi}^
b_L\gamma_{\mu}\psi^a_L)\,+\,({\bar \psi}^a_R\gamma^{\mu}\psi^b_R)
({\bar \psi}^b_R\gamma_{\mu}\psi^a_R)]}~\\~\;\;\;\;\;\;\;\;\;\,=\,-\,
\displaystyle{\frac{4\pi^2G_V}{N_cN_f\Lambda^2_\chi}}[({\bar \psi}\gamma^{\mu}
\psi)({\bar \psi}\gamma_{\mu}\psi)\,+\,({\bar \psi}\gamma_5\gamma^{\mu}\psi)
({\bar \psi}\gamma_5\gamma_{\mu}\psi)]~\\~\;\;\;\;\;\;\;\;\;\;\;\;\;\,-\,
\displaystyle{{\frac {4\pi^2{\hat G}_V}{N_cN_f\Lambda^2_\chi}}\sum_{i=1}^{N^2_f
-1}[({\bar \psi}\gamma^{\mu}t_i\psi)({\bar \psi}\gamma_{\mu}t_i\psi)\,+\,
({\bar \psi}\gamma_5\gamma^{\mu}t_i\psi)({\bar \psi}\gamma_5\gamma_{\mu}t_i
\psi)]}~.
\end{array}
\end{equation}
\noindent
in which $~G_S~$, $~G_V~$ are two coupling constants treated as free
parameters, $~{\hat G}_S\equiv N_fG_S~$, $~{\hat G}_V\equiv N_fG_V~$, and the
flavor group generator $~t_i~$ is normalized as $~t_i=\lambda_i/\sqrt {2}~$,
$~i=1,2,\cdots,N^2_f-1~$. It is argued in Ref. \cite{BBdR} that the NJL type
Lagrangian $~{\cal L}^{SP}_{NJL}\,+\,{\cal L}^{VA}_{NJL}~$ may be understood
as coming from integrating out the high momentum modes of quarks and gluons in
the fundamental theory of QCD and is regarded as the main part in $~{\cal L}_
{QCD}~$ that is responsible for the chiral symmetry breaking, while $~{\cal L}^
{\Lambda_\chi}_{QCD}~$ is responsible for the low-energy gluonic corrections
to the broken chiral symmetry state. This supports the idea of the chiral
quark model \cite{MG}. Even without the gluonic corrections from $~{\cal L}^{
\Lambda_\chi}_{QCD}~$, the NJL type Lagrangian can give rather good
phenomenological predictions for low energy hadron physics \cite{BBdR}.
Since we are not aiming at accurate calculations, we simply neglect the
gluonic corrections in this paper and keep only $~{\cal L}^{SP}_{NJL}\,+\,
{\cal L}^{VA}_{NJL}~$ as the effective interactions between quarks, i.e. the
Lagrangian $~{\cal L}~$ in (4) is taken to be
\begin{eqnarray}           %(17)
{\cal L}\,=\,{\bar \psi}\,i\,/\!\!\! \partial\,\psi\,+\,{\cal L}^
{SP}_{NJL}\,+\,{\cal L}^{VA}_{NJL}\,.
\end{eqnarray}
\noindent
The coefficients $~b_{\Gamma}~$ in (10) and (13) are now
\begin{equation}         %(18)
\begin{array}{ll}
b^1_S\,=\,b^1_P\,=\,\displaystyle{\frac {4\pi^2G_S}{N_cN_f\Lambda^2_\chi}}\,,
\;\;\;\;\;\;\;\;b^{t_i}_S\,=\,b^{t_i}_P\,=\,\displaystyle{\frac{4\pi^2
{\hat G}_S}{N_cN_f\Lambda^2_\chi}}\,,\\
b^1_V\,=\,b^1_A\,=\,-\,\displaystyle{\frac{8\pi^2G_V}{N_cN_f\Lambda^2_\chi}}\,,
\;\;\;\;\;\;\;\;b^{t_i}_V\,=\,b^{t_i}_A\,=\,-\,\displaystyle{\frac {8\pi^2
{\hat G}_V}{N_cN_f\Lambda^2_\chi}}\,,
\end{array}
\end{equation}
\noindent
and the pure source term $~P(K_2,K_4)~$ obtained from (5) is
\begin{eqnarray}          %(19)
P(K_2,K_4)\,=\,\displaystyle{{\frac {1}{2}}\,\sum_{\Gamma} \left(\frac
{(K^{\Gamma}_2)^2}{b^1_{\Gamma}\,+\,b^{t_i}_{\Gamma}\,+\,2K^{\Gamma}_4}\right)}
\,.
\end{eqnarray}

There are three independent free parameters $~G_S~$, $~G_V~$, and
$~\Lambda_\chi~$ in the ENJL model in Ref. \cite{BBdR}. They are related to
the constituent chiral quark mass $~M_Q~$, the quark axial-vector coupling
constant $~g_A~$, and the vector meson mass $~M_V~$ by the following
relations\cite{BBdR}
\begin{eqnarray}               %(20)
1/G_S\,=\,(M_Q/\Lambda_\chi)^2\Gamma(-1,\,(M_Q/\Lambda_\chi)^2)
(1\,+\,\gamma_{-1})\,,
\end{eqnarray}
\begin{eqnarray}               %(21)
\displaystyle{g_A\,=\,\frac{1}{1\,+\,4G_V(M_Q/\Lambda_\chi)^2\Gamma(0,\,
(M_Q/\Lambda_\chi)^2)(1\,+\,\gamma_{01})}}\,,
\end{eqnarray}
\begin{eqnarray}               %(22)
\Lambda_\chi^2\,=\,\frac{2}{3}M_V^2G_V\Gamma(0,\,(M_Q/\Lambda_\chi)^2)(1\,
+\,\gamma_{03})\,,
\end{eqnarray}
\noindent
where $~\Gamma(n-2,\,(M_Q/\Lambda_\chi)^2)~$ is the incomplete gamma function
 \cite{BBdR}, and $~\gamma_{-1}~$, $~\gamma_{01}~$, and $~\gamma_{03}~$ are
gluonic corrections which are actually not large and have been neglected in
our Lagrangian (17). Apart from the gluonic corrections, the rest part of the
formulae used to fit the data in Ref. \cite{BBdR} are all of the leading order
in $~1/N_c~$ expansion. Actually $~M_Q~$, $~\Lambda_\chi~$, and $~g_A~$ are
taken as the input parameters to fit the data in Ref. \cite{BBdR}. There are
five different ways of fitting the data presented in Ref. \cite{BBdR}, which
determine different sets of values of $~M_Q~$, $~\Lambda_\chi~$, and $~g_A~$,
and the predictions are all successful. The simplest one of the fits is their
{\bf Fit 4} in which $~g_A=1~$. From (21) we see that this corresponds to
$~G_V=0~$ which will simplify our calculation. Moreover, it is explained by
Weinberg \cite{W} that $~g_A~$ should actually be unity in the constituent
quark model. In view of this, we take in this paper the set {\bf Fit 4} in
Ref. \cite{BBdR}, and by means of (20)-(22), it leads to
\begin{eqnarray}           %(23)
G_S\,=\,1.19,\;\;\;\;\;\;G_V\,=\,0,\;\;\;\;\;\;\Lambda_\chi\,=\,667\,MeV\,.
\end{eqnarray}

In the present model, $~N_c=N_f=3~$. Let $~N\equiv N_c=N_f~$. Our dynamical
calculation of the condensates is in the $~1/N~$ expasion. In doing so, we
should take a proper limit of $~N\rightarrow\infty~$ such that the three terms
in (17) are of the same order in this limit, i.e. both the propagation of the
quark field and all the NJL type interactions are included. This means that
(cf. (15)-(17)) we should treat the flavor singlet coupling constants $~G_S~$,
$~G_V~$ and the flavor non-singlet coupling constants $~{\hat G}_S~$,$~
{\hat G}_V~$ as independent finite parameters as $~N\rightarrow\infty~$. After
obtaining the condensates to certain order in $~1/N~$ expansion, we then take
numerically $~N=3~$ and $~{\hat G}_S=3G_S~$, $~{\hat G}_V=3G_V~$ when
calculating the enhancement factor $~\kappa~$. In order to make comparison with
 the formulae in Ref. \cite{BBdR}, we define the constituent chiral quark mass
as
\begin{eqnarray}           %(24)
M_Q\,\equiv\,-\,b^1_S\,<{\bar \psi}\psi>\,,
\end{eqnarray}
\noindent
which corresponds to the mean-field approximation interpretation of $M_Q$.
Using the method described above, similar to the calculation presented in Ref.
\cite{HK}, we obtain the following results.

In the large-$N$ limit, $~W[K_2,K_4]~$ is contributed by chain diagrams with
open ends. Then the gap equation (eq.(6) with $~K_2=K_4=0~$) to this order
reads
\begin{equation}          %(25)
\begin{array}{ll}
\displaystyle{<{\bar \psi}\psi>\,=\,-\,i\,tr\,\int_{\Lambda_\chi}
\frac{d^4p}{(2\pi)^4}\,\frac{1}{/\!\!\!p-M_Q}\,=\,-\,\frac{N^2M_Q^3}{4\pi^2}\,
\Gamma(-1,\,(M_Q/\Lambda_\chi)^2)}\;,\\
<{\bar \psi}i\gamma_5\psi>\,=\,<{\bar \psi}\gamma_{\mu}\psi>\,=\,<{\bar
\psi}\gamma_5\gamma_{\mu}\psi>\,=\,0\;.
\end{array}
\end{equation}
\noindent
In (25) the $~\int_{\Lambda_\chi}~$ means that the $~\Lambda_\chi~$-regulator
used in Ref. \cite{BBdR} is taken for the momentum integration. In the chiral
limit $~<{\bar \psi}\psi>=N<{\bar u}u>~$, thus we have
\begin{eqnarray}         %(26)
\displaystyle{<{\bar u}u>\,=\,-\,\frac{N}{4\pi^2}\,M_Q^3\,\Gamma(-1,\,
(M_Q/\Lambda_\chi)^2)}\,.
\end{eqnarray}
\noindent
which coincides with the result in Ref. \cite{BBdR}. It is also easy to see
that, in the large-$N$ limit, $~[{\delta W}/{\delta K^{\Gamma}_4}]_{K_2,K_4=0}
=\Sigma_{\Gamma}^2~$, so that $~[\Xi_{\Gamma}]_{K_2,K_4=0}~$ vanishes. Since in
this limit $~F_{\Gamma}=1~$, $~[\Xi_{\Gamma}]_{K_2,K_4=0}~$ is just the non-
factorized part of the four-quark condensate. Thus we have
\begin{eqnarray}        %(27)
<{\bar \psi}\Gamma\psi{\bar \psi}\Gamma\psi>_{NF}\,=\,0
\end{eqnarray}
\noindent
in the large-$N$ limit. The intuitive reason for the factorization of
the four-quark condensate in the $~N\rightarrow\infty~$ limit is as follows.
Diagrammatically, the four-quark condensate is obtained by removing a
node from the chain diagram for $~W~$, which releases four quark lines.
For open-end chains, the removal of a node causes two disconnected chains
, each with two quark lines, and this is just the factorization of the
four-quark condensate.

To order-$~1/N~$, we should further take into account the P2PI$(\Pi_s)$
diagrams shown in Fig. 6 in Ref. \cite{HK}, which include the closed chain
diagrams. Removing a node from a closed chain does not cause two disconnected
chains, so that factorization breaks down to this order. Up to Next-to-the-
leading order in $~1/N~$ expansion, we see from (3) that only an order-
$1/N~$ calculation of $~<{\bar \psi}\Gamma\psi{\bar \psi}\Gamma\psi>_{NF}~$
 is further needed.

In practical applications, the currents in (1) are usually color octets.
Similar to Ref.\cite{HK}, by applying the method described above, we obtain
the following up to order-$1/N$ results of the non-factorized parts of various
relevant four-quark condensates.
\begin{equation}          %(28)
\begin{array}{ll}
\displaystyle{<{\bar u}\lambda_\alpha u{\bar u}\lambda_\alpha u>_ {NF}\,
 =\,(F_S/N)(S_{NF}\,-\,P_{NF}\,-\,A_{NF})\,+\,O(1/N^2)}\,,\\
\displaystyle{<{\bar u}\lambda_\alpha d{\bar d}\lambda_\alpha u>_ {NF}\,
 =\,(F_S/N)S_{NF}\,+\,O(1/N^2)}\,,\\
\displaystyle{<{\bar u}\lambda_\alpha u{\bar d}\lambda_\alpha d>_{NF}\,
 =\,0\,+\,O(1/N^2)}\,,\\\\
\end{array}
\end{equation}
\begin{equation}        %(29)
\begin{array}{ll}
\displaystyle{<{\bar u}i\gamma_5\lambda_\alpha u{\bar u}i\gamma_5 \lambda_
 \alpha u>_{NF}\,=\,(F_P/N)(S_{NF}\,-\,P_{NF}\,+\,A_{NF})\,+\,
 O(1/N^2)}\,,\\
\displaystyle{<{\bar u}i\gamma_5\lambda_\alpha d{\bar d}i\gamma_5 \lambda_
 \alpha u>_{NF}\,=\,(F_P/N)S_{NF}\,+\,O(1/N^2)}\,,\\
\displaystyle{<{\bar u}i\gamma_5\lambda_\alpha u{\bar d}i\gamma_5 \lambda_
 \alpha d>_{NF}\,=\,0\,+\,O(1/N^2)}\,,\\\\
\end{array}
\end{equation}
\begin{equation}         %(30)
\begin{array}{ll}
\displaystyle{<{\bar u}\gamma^{\mu}\lambda_\alpha u{\bar u}\gamma_{\mu}\lambda_
 \alpha u>_{NF}\,=\,(F_V/N)(S_{NF}\,+\,P_{NF}\,-\,\frac{1}{2}A_{NF})\,+
 \,O(1/N^2)}\,,\\
\displaystyle{<{\bar u}\gamma^{\mu}\lambda_\alpha d{\bar d}\gamma_{\mu}\lambda_
 \alpha u>_{NF}\,=\,(F_V/N)S_{NF}\,+\,O(1/N^2)}\,,\\
\displaystyle{<{\bar u}\gamma^{\mu}\lambda_\alpha u{\bar d}\gamma_{\mu}\lambda_
 \alpha d>_{NF}\,=\,0\,+\,O(1/N^2)}\,,\\\\
\end{array}
\end{equation}
\begin{equation}         %(31)
\begin{array}{ll}
\displaystyle{<{\bar u}\gamma_5\gamma^{\mu}\lambda_\alpha u{\bar u}\gamma_5
 \gamma_{\mu}\lambda_\alpha u>_{NF}\,=\,(F_A/N)(S_{NF}\,+\,P_{NF}\,+\,
 \frac{1}{2}A_{NF})\,+\,O(1/N^2)}\,,\\
\displaystyle{<{\bar u}\gamma_5\gamma^{\mu}\lambda_\alpha d{\bar d}\gamma_5
 \gamma_{\mu}\lambda_\alpha u>_{NF}\,=\,\frac{F_A}{N}S_{NF}\,+\, O(1/N^2)}\,,\\
\displaystyle{<{\bar u}\gamma_5\gamma^{\mu}\lambda_\alpha u{\bar d}\gamma_5
 \gamma_{\mu}\lambda_\alpha d>_{NF}\,=\,0\,+\,O(1/N^2)}\,,\\\\
\end{array}
\end{equation}
\begin{equation}        %(32)
\begin{array}{ll}
\displaystyle{<{\bar u}\sigma^{\mu\nu}\lambda_\alpha u{\bar u}\sigma_{\mu\nu}
 \lambda_\alpha u>_{NF}\,=\,(F_T/N)(S_{NF}\,-\,P_{NF})\,+\,O(1/N^2)}\,,\\
\displaystyle{<{\bar u}\sigma^{\mu\nu}\lambda_\alpha d{\bar d}\sigma_{\mu\nu}
 \lambda_\alpha u>_{NF}\,=\,(F_T/N)S_{NF}\,+\,O(1/N^2)}\,,\\
\displaystyle{<{\bar u}\sigma^{\mu\nu}\lambda_\alpha u{\bar d}\sigma_{\mu\nu}
 \lambda_\alpha d>_{NF}\,=\,0\,+\,O(1/N^2)}\,,\\\\
\end{array}
\end{equation}
\noindent
where
\begin{equation}        %(33)
\begin{array}{ll}
\displaystyle{F_S\,=\,-\,F_P\,=\,-\,\frac{1}{2}\,+\,O(1/N^2)}\,,\\
\displaystyle{F_V\,=\,-\,F_A\,=\,-2\,+\,O(1/N^2)}\,,\\
\displaystyle{F_T\,=\,-\,6\,+O(1/N^2)}\,,
\end{array}
\end{equation}
\noindent
and
\begin{equation}          %(34)
\begin{array}{ll}
S_{NF}\,=\,-\,\displaystyle{\frac{i}{4}\frac{\pi^2G_S}{N\Lambda_\chi^2}\int_
{\Lambda_\chi}\frac{d^4q}{(2\pi)^4}\frac{T^2_S(q^2,M_Q)}{1-\frac{\pi^2G_S}
{N\Lambda_\chi^2}T_S(q^2,M_Q)}}\,,\\\\
P_{NF}\,=\,-\,\displaystyle{\frac{i}{4}\frac{\pi^2G_S}{N\Lambda_\chi^2}\int_
{\Lambda_\chi}\frac{d^4q}{(2\pi)^4}\frac{T^2_P(q^2,M_Q)}{1-\frac{\pi^2G_S}
{N\Lambda_\chi^2}T_P(q^2,M_Q)}}\,,\\\\
A_{NF}\,=\,-\,\displaystyle{\frac{i}{4}\frac{\pi^2G_S}{N\Lambda_\chi^2}\int_
{\Lambda_\chi}\frac{d^4q}{(2\pi)^4}\frac{q^{-2}T_{PA}(q^2,M_Q)
T_{AP}(q^2,M_Q)}{1-\frac{\pi^2G_S}{N\Lambda_\chi^2}T_P(q^2,M_Q)}}\,,
\end{array}
\end{equation}
\noindent
in which
\begin{equation}         %(35)
\begin{array}{ll}
T_S(q^2,M_Q)\,=\,\displaystyle{4iN\int_{\Lambda_\chi} \frac{d^4p}{(2\pi)^4}\,
tr\,[\frac{1}{/\!\!\!p+/\!\!\!q/2-M_Q}\frac{1}{/\!\!\!p-/\!\!\!q/2-M_Q}]}\\
\,\;\;\;\;\;\;\;\;\;\;\;\;\;\;\;\;\;\;\;=\,\displaystyle{\frac{NM_Q^2}{\pi^2}\,
\Gamma(-1,\,(\frac{M_Q}{\Lambda_\chi})^2)}\,-\,(2q^2-8M_Q^2)\,I(q^2,\,M_Q)\,,\\
T_P(q^2,M_Q)\,=\,\displaystyle{4iN\int_{\Lambda_\chi} \frac{d^4p}{(2\pi)^4}\,
tr\,[i\gamma_5\frac{1}{/\!\!\!p+/\!\!\!q/2-M_Q}i\gamma_5\frac{1}{/\!\!\!p-
/\!\!\!q/2-M_Q}]}\\
\,\;\;\;\;\;\;\;\;\;\;\;\;\;\;\;\;\;\;\;=\,\displaystyle{\frac{NM_Q^2}{\pi^2}\,
\Gamma(-1,\,(\frac{M_Q}{\Lambda_\chi})^2)}\,-\,2q^2\,I(q^2,\,M_Q)\,,\\
T_{PA}(q^2,M_Q)\,=-T_{AP}(q^2,M_Q)\,=\,\displaystyle{4iN\int_{\Lambda_\chi}
 \frac{d^4p}{(2\pi)^4}\,tr\,[i\gamma_5\frac{1}{/\!\!\!p+/\!\!\!q/2-M_Q}
\gamma_5{/\!\!\!q}\frac{1}{/\!\!\!p-/\!\!\!q/2-M_Q}]}\\
\,\;\;\;\;\;\;\;\;\;\;\;\;\;\;\;\;\;\;\;=\,\displaystyle{-\,i\,M_Q\,q^2\,
I(q^2,\,M_Q)}\,,\\
I(q^2,M_Q)\,\equiv\,\displaystyle{-\,\frac{N}{4\pi^2}\int^1_0dx\,\Gamma(0,\,
\frac{q^2(x^2-x)+M_Q^2}{\Lambda^2_\chi})}\,.\\
\end{array}
\end{equation}
\noindent
In (35) $~tr~$ is the trace in the Dirac spinor space. With all these,
we can calculate the enhancement factor $~\kappa~$ for practical processes.

Take the $~\rho$-meson sum rule as an example. The relevant four-quark
condensates are \cite{LNT}
\begin{equation}            %(36)
\begin{array}{ll}
C_6<O_6>\,=\,6\pi^3\alpha_s\{<({\bar u}\gamma^{\mu}\gamma_5\lambda_{\alpha}u
-{\bar d}\gamma^{\mu}\gamma_5\lambda_{\alpha}d)^2>\\\;\;\;\;\;\;\;\;\;\;\;\;\;
\;\;\;\;\;+\displaystyle{\frac{2}{9}
<({\bar u}\gamma^{\mu}\lambda_{\alpha}u+{\bar d}\gamma^{\mu}\lambda_{\alpha}
d)\sum_{u,d,s}{\bar \psi}\gamma_{\mu}\lambda_{\alpha}\psi>\}}.
\end{array}
\end{equation}
\noindent
Substituting (36) into (3), and using the results in (30) and (31), we obtain
the enhancement factor for the $~\rho$-meson sum rule
\begin{equation}           %(37)
\begin{array}{ll}
\kappa\,=\,1\,+\,\displaystyle{\frac{1}{3}\,\frac{S_{NF}+P_{NF}}{<{\bar u}u>
^2}\,+\,\frac{11}{42}\,\frac{A_{NF}}{<{\bar u}u>^2}\,+\,O(\frac{1}{N^2})}\,.
\end{array}
\end{equation}

When doing numerical calculations, we should bear in mind that the numbers in
(23) are obtained from fitting the data with the large-$N_c$ formulae together
with gluonic corrections, while our present calculation is up to order-$1/N$
without gluonic corrections. Thus we should not take the numbers in (23) so
seriously when calculating the value of $~\kappa~$. From (36) we see that the
values of $~S_{NF}, \,P_{NF}~$ and $~A_{NF}~$ are only related to the two
parameters $~G_S~$ and $~M_Q^2/\Lambda_\chi^2~$ which are further related to
each other by (20). Therefore our $~\kappa~$ is a function of only one
parameter $~G_S~$. In {\bf Table 1}, we list the values of $~S_{NF},\,P_{NF}~$
and $~A_{NF}~$ for several values of $~G_S~$ around the number given in (23).
The corresponding numbers of $~\kappa~$ in the $~\rho$-meson sum rule are
listed in {\bf Table 2}. We see that in the small $~G_S~$ region $~\kappa~$
is a sensitive function of $~G_S~$, and $~\kappa=2.4~$ is included in this
region. At present $~G_S~$ is not determined accuratly, so that we still cannot
 predict quantitatively if $~\kappa~$ is precisely 2.4. However, due to this
sensitivity, the possibility of having $~\kappa=2.4~$ mainly depends on the
positivity of the order-$1/N~$ contribution in (37) (the sum of the second and
third terms in (37)) which is not obvious in general. Our calculation shows
that it is really the case.

 Inview of the fact that $~\kappa~$ is sensitive to the value of $~G_S~$ which
is not yet well determined, our conclusion in this paper is that {\it the
explanation of $~\kappa=2.4~$ as the contributions of the non-factorized
parts of the four-quark condensates is possible}.

\end{sf}

\newpage

\vspace{3cm}
\begin{center}
{\bf Table 1}   Values of $~S_{NF}~$, $~P_{NF}~$, and $~A_{NF}~$ ( in the
unit of $~<{\bar u}u>^2~$) for given values of $~G_S~$.\\
\vspace{0.5cm}
\normalsize
\begin{tabular}{*{8}{c}}
      \hline
      \hline
\multicolumn{1} {l}{$G_{s}$} & {2.0}
 &\multicolumn{1} {c}{1.5} & {1.3}
 &\multicolumn{1} {c}{1.25} & {1.2}
 &\multicolumn{1} {c}{1.15} & {1.1}\\
   \hline
\multicolumn{1} {l}{$S_{NF}$} & {0.00069}
 &\multicolumn{1} {c}{0.0028} & {0.0079}
 &\multicolumn{1} {c}{0.011} & {0.016}
 &\multicolumn{1} {c}{0.026} & {0.047}\\
   \hline 
\multicolumn{1} {l}{$P_{NF}$} & {0.010}
 &\multicolumn{1} {c}{0.017} & {0.026}
 &\multicolumn{1} {c}{0.030} & {0.037}
 &\multicolumn{1} {c}{0.048} & {0.072}\\
      \hline
\multicolumn{1} {l}{$A_{NF}$} & {-0.00054}
 &\multicolumn{1} {c}{-0.00077} & {-0.00094}
 &\multicolumn{1} {c}{-0.00099} & {-0.0011}
 &\multicolumn{1} {c}{-0.0011} & {-0.0012}\\
      \hline
\\
      \hline
\multicolumn{1} {l}{$G_{s}$} & {1.08}
 &\multicolumn{1} {c}{1.05} & {1.05}
 &\multicolumn{1} {c}{1.02} & {1.01}
 &\multicolumn{1} {c}{1.009} & {1.008}\\
   \hline
\multicolumn{1} {l}{$S_{NF}$} & {0.065}
 &\multicolumn{1} {c}{0.12} & {0.23}
 &\multicolumn{1} {c}{0.37} & {0.85}
 &\multicolumn{1} {c}{0.96} & {1.1}\\
   \hline
\multicolumn{1} {l}{$P_{NF}$} & {0.090}
 &\multicolumn{1} {c}{0.15} & {0.26}
 &\multicolumn{1} {c}{0.40} & {0.88}
 &\multicolumn{1} {c}{0.99} & {1.1}\\
      \hline
\multicolumn{1} {l}{$A_{NF}$} & {-0.0012}
 &\multicolumn{1} {c}{-0.0013} & {-0.0013}
 &\multicolumn{1} {c}{-0.0014} & {-0.0014}
 &\multicolumn{1} {c}{-0.0014} & {-0.0014}\\
      \hline
\\
      \hline
\multicolumn{1} {l}{$G_{s}$} & {1.007}
 &\multicolumn{1} {c}{1.006} & {1.005}
 &\multicolumn{1} {c}{1.004} & {1.003}
 &\multicolumn{1} {c}{1.002} & {1.001}\\
   \hline
\multicolumn{1} {l}{$S_{NF}$} & {1.3}
 &\multicolumn{1} {c}{1.5} & {1.9}
 &\multicolumn{1} {c}{2.4} & {3.4}
 &\multicolumn{1} {c}{5.3} & {12}\\
   \hline
\multicolumn{1} {l}{$P_{NF}$} & {1.3}
 &\multicolumn{1} {c}{1.6} & {1.9}
 &\multicolumn{1} {c}{2.5} & {3.4}
 &\multicolumn{1} {c}{5.3} & {12}\\
      \hline
\multicolumn{1} {l}{$A_{NF}$} & {-0.0014}
 &\multicolumn{1} {c}{-0.0014} & {-0.0014}
 &\multicolumn{1} {c}{-0.0014} & {-0.0014}
 &\multicolumn{1} {c}{-0.0014} & {-0.0014}\\
      \hline
      \hline
\end{tabular}
\end{center}

\vspace{1cm}
\begin{center}
\noindent
{\bf Table 2}   Values of the enhancement factor $~\kappa~$ ( eq.(37)) for
given values of $~G_S~$.\\
\vspace{0.5cm}

\normalsize
\begin{tabular}{*{8}{c}}
      \hline
      \hline
\multicolumn{1} {l}{$G_{s}$} & {2.0}
 &\multicolumn{1} {c}{1.5} & {1.3}
 &\multicolumn{1} {c}{1.25} & {1.2}
 &\multicolumn{1} {c}{1.15} & {1.1}\\
   \hline
\multicolumn{1} {l}{$\kappa$} & {1.0035}
 &\multicolumn{1} {c}{1.0065} & {1.011}
 &\multicolumn{1} {c}{1.014} & {1.017}
 &\multicolumn{1} {c}{1.024} & {1.039}\\
   \hline
\\
   \hline
\multicolumn{1} {l}{$G_{s}$} & {1.08}
 &\multicolumn{1} {c}{1.05} & {1.03}
 &\multicolumn{1} {c}{1.02} & {1.01}
 &\multicolumn{1} {c}{1.009} & {1.008}\\
   \hline
\multicolumn{1} {l}{$\kappa$} & {1.051}
 &\multicolumn{1} {c}{1.089} & {1.16}
 &\multicolumn{1} {c}{1.26} & {1.57}
 &\multicolumn{1} {c}{1.65} & {1.74}\\
   \hline
\\
   \hline
\multicolumn{1} {l}{$G_{s}$} & {1.007}
 &\multicolumn{1} {c}{1.006} & {1.005}
 &\multicolumn{1} {c}{1.004} & {1.003}
 &\multicolumn{1} {c}{1.002} & {1.001}\\
   \hline
\multicolumn{1} {l}{$\kappa$} & {1.86}
 &\multicolumn{1} {c}{2.0} & {2.3}
 &\multicolumn{1} {c}{2.6} & {3.3}
 &\multicolumn{1} {c}{4.6} & {8.7}\\
   \hline
   \hline
\end{tabular}
\end{center}

\end{document}